\documentclass[preprint,aps]{revtex4}

\usepackage{graphicx}
\usepackage{dcolumn}
\usepackage{bm}

\begin{document}
\preprint{Stanford, Bethlehem and Frankfurt am main}
\title{Charmonium dissociation in the context of Bethe-Salpeter
equation at finite temperature}

\author{I. Zakout}
\affiliation{%
Physics Department, Stanford University, 
Stanford, CA 94305-4060}
\author{H. R. Jaqaman}
\affiliation{%
Physics Department, Bethlehem University, P.O. Box 9, Bethlehem, 
Palestinian Authority}

\author{W. Greiner}
\affiliation{
Institut f\"ur Theoretische Physik, Robert Mayer Str. 8-10,
D-60054, Frankfurt am Main, Germany}

\begin{abstract}
We derive the spin-dependent Bethe-Salpeter equation for
$q\overline{q}$ bound state at finite temperature
in the ladder and cross box diagram approximations.
The imaginary time formalism is adopted in the adiabatic approximation where the flux tube thermal excitation is separated from the thermal excitation of the constituent quarks. 
The thermal excitation of the flux tube is implemented by using a temperature dependent potential inferred from the lattice gauge calculations. 
The thermal excitations of the constituent quarks are evaluated by summing Matsubara frequencies. 
The resultant equation is used to study the thermal charmonium spontaneous dissociation mechanism at temperatures below the critical one of the phase transition to the quark gluon plasma. 
Our results show that when the thermal excitation of the constituent quarks is considered beside the flux tube thermal excitation, the spontaneous dissociation unlikely takes place in the hadronic phase. 
In contrary, when the thermal excitation of the flux tube is considered only in the calculations, the spontaneous dissociation likely takes place just below the critical temperature.
\end{abstract}


\maketitle
\section{Introduction}
Charmonium suppression/enhancement in the heavy ions collision experiments has received much attention in the search of the quark gluon plasma (QGP)
\cite{Barnes1,Barnes2,Wong1,Satz1,Satz2,Rapp1,Sibirtsev1}.
Whatever signal is chosen for the identification of QGP, contributions to that signal from conventional hadronic processes must be identified as backgrounds and removed from the data. 
Among of these signals is the signal from the charmoinum dissociation at temperatures below the QGP phase transition temperature $T_c$.
There is a considerable uncertainty on the origin of the anomalous suppression due to the lack of a reliable information on $J/\psi$ and $\chi_J$ dissociation below $T_c$.
It is known that $J/\psi$ production in hadron-hadron collisions is to 
a considerable extent due to the production and subsequent decay of higher excited $c\overline{c}$ states. 
Since different quarkonium states have different sizes, one expects that higher excited states will dissolve at smaller temperatures than the more tightly bound ground states\cite{Satz1,Satz2}. 
These facts may lead to a sequential suppression pattern in $J/\psi$ yield in nucleus-nucleus collisions as a function of the energy density\cite{Barnes1,Barnes2}.
The charmonium can dissociate spontaneously when its mass spectrum becomes unbound and equal to a pair of final open charm mesons at temperatures below the critical temperature $T_c$.

The spontaneous dissociation has been studied by using a temperature-dependent potential inferred from the lattice gauge calculations\cite{Kaczmarek1}.  
In consideration of the spontaneous dissociation below $T_c$, it is necessary to find the selection rules for the dissociation of a heavy quarkonium state with initial quantum numbers $J, L_i$ and $S_i$ into a pair of open charm mesons with a total spin $S$ and a relative orbital angular momentum $L$. 
The total spin $S$ is conserved. Parity conservation requires $\Delta=|L-L_i|=1$ in this dissociation.
Hence $J/\psi$ and $\psi'$ dissociates into a pair of open charm mesons with $L=1$, while $\chi$ dissociates into a pair of open charm mesons with $L=0$ or $2$. 
These spin and angular momentum quantum numbers give rise to the selection rules for final meson states and alter the dissociation threshold energies and dissociation temperatures\cite{Barnes1,Barnes2}.
The spontaneous thermal dissociation also takes place when the charmonium absorbs a light meson such as a pion and their total mass becomes equal to a pair of open charm mesons via the threshold production interaction
\begin{eqnarray}
m_{J/\psi}+m_{\pi}=m_{D\overline{D}}.
\end{eqnarray}
The selection rules still play a significant rule in this interaction.
However, in the hadronic phase the pion gas is accumulated in the media 
and it is likely that charmonium dissociates thermally spontaneously  by absorbing a pion.

The other possibility is that the quarkonium can dissociate by colliding with light hadrons\cite{Barnes1,Barnes2,Sibirtsev1,Muller1,Blaschke1}. 
This co-mover dissociation process must be understood and incorporated 
in the simulation of heavy ion collisions before the QGP formation can be established through this signature.
The nonrelativistic quark-interchange model has been used to evaluate the 
low-energy cross sections of heavy quarkonium in collision with
light mesons such as $\pi, \rho$ and $K$ in terms of wave functions and interactions at the quark level.
It is found that as the temperature increases, the threshold energy decreases and the dissociation cross section increases\cite{Barnes1,Barnes2}.
Furthermore, any quantitative analysis of $J/\psi$ dissociation in 
nucleus-nucleus collision should include the effects of the meson mass modification in the dense and hot matter.

The relativistic quasi-potential at finite temperature for the $q\overline{q}$ is essential to study the charm spectroscopy at finite temperature. 
Furthermore, it is expected that the relativistic corrections for the wave function shall be significant and they may change the results of the co-mover dissociation processes.
The inter-quark temperature dependent potential modifies the spectrum significantly. 
Altogether, the thermal kinetic excitations of the constituent quarks are 
presumably supposed to change the full BS equation results.

The outline of the paper is as follows. In Sec. II, we shall derive the BS equation at finite temperature. The adiabatic approximation is considered where the flux tube thermal excitation is separated from the kinetic thermal excitation of the constituent quarks.
The interaction of the constituent quarks with the mean fields is also considered.
In Sec. III, we study the thermal spontaneous dissociation at $T<T_c$ and give our conclusions.

\section{Bethe-Salpeter in hot and dense medium}

The Bethe-Salpeter equation in the ladder approximation for the quark ($1$) and
anti-quark ($\overline{2}$) reads
\begin{eqnarray}
{\cal G}^{-1}_1(p_1)\Psi(p)\overline{\cal G}^{-1}_2(p_2)
=i\int \frac{d^4 p}{(2\pi)^4} {\cal K}(p,p')\Psi(p').
\end{eqnarray}  
The quark propagator can be calculated by solving the Schwinger-Dyson equation in the ladder approximation. However, in the heavy-heavy and heavy-light meson bound states, the self-energy corrections are small. Therefore, we consider the free quark propagator in the present calculations.
When the meson is probed in the nuclear matter, the constituent quark interacts with the medium by coupling the scalar and vector mean fields.
In the mean field approximations, the bound state equation probed in the nuclear matter becomes
\begin{eqnarray}
[p_{1\mu}\gamma^{\mu}-m_1+ (S^{(1)}-V^{(1)}_{\mu}\gamma^{\mu})]
\Psi(p)[p_{2\mu}\gamma^{\mu}+m_2+
(S^{(\overline{2})}-V^{(\overline{2})}_{\mu}\gamma^{\mu})]
=i\int \frac{d^4 p}{(2\pi)^4} {\cal K}(p,p')\Psi(p').
\end{eqnarray}

In symmetric isotropic nuclear matter, the vector field is simplified to time-like vector field,
\begin{eqnarray}
V^{(i)}_{\mu}\gamma^{\mu}\approx V^{(i)}\beta.
\end{eqnarray}
The bound state equation is simplified by introducing the following parameters,
\begin{eqnarray}
M_1&=&\eta_1 M^{*}-V^{(1)} \nonumber\\
M_2&=&\eta_2 M^{*}-V^{(\overline{2})} \nonumber\\  
M_{12}&=&M^{*}-[V^{(1)}+V^{(\overline{2})}].
\end{eqnarray}
The interaction with the vector mean fields can be absorbed in this new parameterization. Hence, the vector mean fields, effectively, couple with the bound state mass.
This simplifies the equation drastically since the coupling with the vector fields will not appear any more.
On the other hand, the scalar mean fields couple with the constituent quarks and modify their masses as follows,
\begin{eqnarray}
m^*_1=m_1-S^{(1)}\nonumber\\
m^*_2=m_2+S^{(\overline{2})}.
\end{eqnarray}

The BS equation probed in the nuclear matter reads,
\begin{eqnarray}
\Psi(p)=
\left[\frac{(M_1+p_0)\beta-{\bf p}\cdot{\bf \gamma}+m^*_1}
{(M_1+p_0)^2-{\epsilon^*_1}^2({\bf p})+ i\delta}\right]
i\int \frac{d^4 p}{(2\pi)^4} {\cal K}(p,p')\Psi(p')
\left[\frac{(p_0-M_2)\beta-{\bf p}\cdot{\bf \gamma}+m^*_2}
{{\epsilon^*_2}^2({\bf p})-(M_2-p_0)^2-i\delta}\right],
\end{eqnarray}
where the effective constituent quark energy reads
\begin{eqnarray}
\epsilon^*_i({\bf p})=\sqrt{{\bf p}^2+{m^*_i}^2}.
\end{eqnarray}
The quasipotential equation in the nuclear matter has the same criteria of the normal quark-meson coupling model.
Hence, the interaction with the mean scalar fields modifies the quark effective masses while the interaction with the vector mean fields modifies the effective bound state mass $M^{*}$.

When the bound state equation is embedded in the thermal bath, the constituent quarks are excited thermally as well as the flux tube. 
To simplify the equation, the thermal excitation of the flux tube between the constituent quarks is separated from the quark kinetic thermal excitations.
In the adiabatic approximation, the flux tube is excited with the temperature independently of the thermal excitations of constituent quarks.
Therefore, the flux tube excitation is not modified with the thermal excitation of the quark propagator.
We assume the instantaneous interaction approximation where the potential doesn't depend on the time-like component.
Therefore, the kernel is approximated as follows,
\begin{eqnarray}
{\cal K}(p,p')\approx {\cal V}({\bf p}-{\bf p}',T,\mu_g).
\end{eqnarray}
The four-dimensional kernel integration is reduced to a three-dimensional one in the term of the instantaneous wave function
$\phi({\bf p})$
\begin{eqnarray}
i\int \frac{d^4 p'}{(2\pi)^4}{\cal K}(p,p')\Psi(p')
=i\int \frac{d^3 {\bf p}'}{(2\pi)^3}
{\cal V}({\bf p}-{\bf p}',T,\mu_g)\phi({\bf p}'),
\end{eqnarray}
where the bound state wave function in the adiabatic and instantaneous  approximations reads
\begin{eqnarray}
\phi({\bf p})=\int \frac{d p_0}{2\pi} \Psi(p).
\end{eqnarray}

In order to simplify the equation, we define the convention
\begin{eqnarray}
{\cal V}(T,\mu_g)\phi
=\int \frac{d^3 {\bf p}'}{(2\pi)^3}{\cal V}({\bf p}-{\bf p}',T,\mu_g)
\phi({\bf p}').
\end{eqnarray}
The three-dimensional bound state equation with the ladder graphs reads,
\begin{eqnarray}
\phi({\bf p})=i\int \frac{d p_0}{2\pi}
\frac{(M_1+p_0)\beta-{\bf \gamma}\cdot{\bf p}+m^*_1}
{(M_1+p_0)^2-{\epsilon^*_1}^2({\bf p})+i\delta}
{\cal V}(T,\mu_g)\phi
\frac{(p_0-M_2)\beta-{\bf \gamma}\cdot{\bf p}+m^*_2}
{{\epsilon^*_2}^2({\bf p})-(p_0-M_2)^2-i\delta}.
\end{eqnarray}
The contribution of the cross box diagram justified by using the Eikonal approximation is added in the equation\cite{Wallace1,Mandelzweig1}.
It is given by
\begin{eqnarray}
I_{\mbox{Cross}}=
i\int \frac{d p_0}{2\pi}
\frac{(M_1+p_0)\beta-{\bf \gamma}\cdot{\bf p}+m^*_1}
{(M_1+p_0)^2-{\epsilon^*_1}^2({\bf p})+i\delta}
{\cal V}(T,\mu_g)\phi
\frac{(p_0+M_2)\beta-{\bf \gamma}\cdot{\bf p}+m^*_2}
{{\epsilon^*_2}^2({\bf p})-(p_0+M_2)^2-i\delta}.
\end{eqnarray}
This approximation is useful since it involves the same argument for both the ladder and cross-box graphs.
It is noted that this term doesn't alter the invariance.
This equation is symmetric in the particle labels and it reduces to the one body Dirac equation when either particle's mass becomes infinite.

For the bound state problem in the center of mass frame $M_{12}=M_{1}+M_{2}$
is the eigenvalue while the subsidiary condition
$M_1-M_2=(m^2_1-m^2_2)/M_{12}$ should be used to define the energy difference. 
This condition guarantees that the correct nonrelativistic limit is used.  
The ladder and cross box diagrams are decomposed as follows,
\begin{eqnarray}
I_{\mbox{Ladder}}=
I^{(0)}_{\mbox{Ladder}}+I^{(1)}_{\mbox{Ladder}}
\end{eqnarray}   
and
\begin{eqnarray}
I_{\mbox{Cross}}=
I^{(0)}_{\mbox{Cross}}+I^{(1)}_{\mbox{Cross}}.
\end{eqnarray}
The BS equation is simplified drastically when it is projected into the plane wave components.
The positive and negative energy projectors read
\begin{eqnarray}
\Lambda^{*i}_{\pm}({\bf p})=
\frac{\epsilon^*_i({\bf p})\pm\beta({\bf\gamma}\cdot{\bf p}+m^*_i)}
{2\epsilon^*_i({\bf p})}.
\end{eqnarray}

For the Ladder graphs we have
\begin{eqnarray}
I^{(0)}_{\mbox{Ladder}}=I^{+-}_{\mbox{Ladder}}+I^{-+}_{\mbox{Ladder}},
\end{eqnarray}
where
\begin{eqnarray}
I^{(+-)}_{\mbox{Ladder}}=
-i\int \frac{dp_0}{2\pi} \left[
\frac{\Lambda^{*(1)}_{+}({\bf p})}
{(p_0+M_1-\epsilon^*_1({\bf p})+i\delta)}
\beta {\cal V}(T,\mu_g)\phi\beta
\frac{\Lambda^{*(2)}_{-}(-{\bf p})}
{(p_0-M_2+\epsilon^*_2({\bf p})-i\delta)}
\right]
\end{eqnarray}
\begin{eqnarray}
I^{(-+)}_{\mbox{Ladder}}=
-i\int \frac{dp_0}{2\pi} \left[
\frac{\Lambda^{*(1)}_{-}({\bf p})}
{(p_0+M_1+\epsilon^*_1({\bf p})-i\delta)}
\beta {\cal V}(T,\mu_g)\phi({\bf p})\beta
\frac{\Lambda^{*(2)}_{+}(-{\bf p})}
{(p_0-M_2-\epsilon^*_2({\bf p})+i\delta)}
\right].
\end{eqnarray}  
These terms lead to the Salpeter equation at zero temperature.
The additional components in the ladder graphs read
\begin{eqnarray}
I^{(1)}_{\mbox{Ladder}}=I^{++}_{\mbox{Ladder}}+I^{--}_{\mbox{Ladder}},
\end{eqnarray}
where
\begin{eqnarray}
I^{(++)}_{\mbox{Ladder}}=
-i\int \frac{dp_0}{2\pi} \left[
\frac{\Lambda^{*(1)}_{+}({\bf p})}
{(p_0+M_1-\epsilon^*_1({\bf p})+i\delta)}
\beta {\cal V}(T,\mu_g)\phi\beta
\frac{\Lambda^{*(2)}_{+}(-{\bf p})}
{(p_0-M_2-\epsilon^*_2({\bf p})+i\delta)}
\right]
\end{eqnarray}
\begin{eqnarray}
I^{(--)}_{\mbox{Ladder}}=
-i\int \frac{dp_0}{2\pi} \left[
\frac{\Lambda^{*(1)}_{-}({\bf p})}
{(p_0+M_1+\epsilon^*_1({\bf p})-i\delta)}
\beta {\cal V}(T,\mu_g)\phi\beta
\frac{\Lambda^{*(2)}_{-}(-{\bf p})}
{(p_0-M_2+\epsilon^*_2({\bf p})-i\delta)}
\right]
\end{eqnarray}  
These components vanish in the zero temperature case.
The $"++"$ and $"--"$ which are essential to obtaining the Dirac
limit are absent at zero temperature unless the cross-box gluon exchange term is included.
These terms correspond to the $Z$-graph contributions of time ordered
perturbation theory.
The contribution of the cross graphs read
\begin{eqnarray}
I^{(0)}_{\mbox{Cross}}=
I^{++}_{\mbox{Cross}} + I^{--}_{\mbox{Cross}},
\end{eqnarray}  
where
\begin{eqnarray}
I^{(++)}_{\mbox{Cross}}=
i\int \frac{dp_0}{2\pi} \left[
\frac{\Lambda^{*(1)}_{+}({\bf p})}
{(p_0+M_1-\epsilon^*_1({\bf p})+i\delta)}
\beta {\cal V}(T,\mu_g)\phi\beta   
\frac{\Lambda^{*(2)}_{+}(-{\bf p})}
{(p_0+M_2+\epsilon^*_2({\bf p})-i\delta)}
\right],
\end{eqnarray}
\begin{eqnarray}
I^{(--)}_{\mbox{Cross}}=
i\int \frac{dp_0}{2\pi} \left[
\frac{\Lambda^{*(1)}_{-}({\bf p})}
{(p_0+M_1+\epsilon^*_1({\bf p})-i\delta)}
\beta {\cal V}(T,\mu_g)\phi\beta
\frac{\Lambda^{*(2)}_{-}(-{\bf p})}
{(p_0+M_2-\epsilon^*_2({\bf p})+i\delta)}
\right].
\end{eqnarray}

The additional terms read
\begin{eqnarray}
I^{(1)}_{\mbox{Cross}}=
I^{+-}_{\mbox{Cross}} + I^{-+}_{\mbox{Cross}},
\end{eqnarray}
\begin{eqnarray}
I^{(+-)}_{\mbox{Cross}}=
i\int \frac{dp_0}{2\pi} \left[
\frac{\Lambda^{*(1)}_{+}({\bf p})} 
{(p_0+M_1-\epsilon^*_1({\bf p})+i\delta)}
\beta {\cal V}(T,\mu_g)\phi\beta
\frac{\Lambda^{*(2)}_{-}(-{\bf p})}
{(p_0+M_2-\epsilon^*_2({\bf p})+i\delta)}
\right],
\end{eqnarray}
\begin{eqnarray}
I^{(-+)}_{\mbox{Cross}}=
i\int \frac{dp_0}{2\pi} \left[
\frac{\Lambda^{*(1)}_{-}({\bf p})}
{(p_0+M_1+\epsilon^*_1({\bf p})-i\delta)}
\beta {\cal V}(T,\mu_g)\phi({\bf p})\beta
\frac{\Lambda^{*(2)}_{+}(-{\bf p})}
{(p_0+M_2+\epsilon^*_2({\bf p})-i\delta)}
\right].
\end{eqnarray}  
These terms vanish at zero temperature by direct integration 
over residues.

The bound state equation reads
\begin{eqnarray}
\phi({\bf p})=I^{(0)}_{\mbox{Ladder}}+I^{(0)}_{\mbox{Cross box}}
+\delta V^{(1)}_{\mbox{T}},
\end{eqnarray}
where
\begin{eqnarray}
\delta V^{(1)}_{\mbox{T}}=I^{(1)}_{\mbox{Ladder}}+I^{(1)}_{\mbox{Cross}}.
\end{eqnarray}
Using the imaginary time formalism preforms the thermal extension.
The right hand side is expanded with respect to the Matsubara frequencies
\begin{eqnarray}
\phi({\bf p})&=&\int^{\infty}_{-\infty}
\frac{dp_0}{2\pi} \left[
{\cal I}^{(0)}_{\mbox{Ladder}}
+{\cal I}^{(0)}_{\mbox{Cross}} \right] + \delta V^{(1)}_{\mbox{T}}
\nonumber\\
&=&iT\sum_{n=-\infty}^{\infty}\left[
{\cal I}^{(0)}_{\mbox{Ladder}}
+{\cal I}^{(0)}_{\mbox{Cross}} \right]_{p_0=i\omega_n} + \delta V^{(1)}_{\mbox{T}}
\nonumber\\     
&=& -i\frac{1}{2}\sum \mbox{Res}\left[
{\cal I}^{(0)}_{\mbox{Ladder}}
+{\cal I}^{(0)}_{\mbox{Cross}} \right]_{p_0}\tanh(p_0/2T) + \delta V^{(1)}_{\mbox{T}},
\end{eqnarray}
where $\omega_n=\pi(2n+1)T$.

The quasipotential equation at finite temperature reads
\begin{eqnarray}
\phi({\bf p})&=&
\left[
{\cal N}_{+-}(T)\frac{
\Lambda^{(1)*}_{+}
({\bf p})\beta{\cal V}(T,\mu_g)\phi\beta
\Lambda^{(2)*}_{-}(-{\bf p})}
{M_{12}-\epsilon^*_1({\bf p})-\epsilon^*_2({\bf p})} \right.
\nonumber\\
&+&
{\cal N}_{-+}(T)\frac{
\Lambda^{(1)*}_{-}
({\bf p})\beta{\cal V}(T,\mu_g)
\phi\beta\Lambda^{(2)*}_{+}(-{\bf p})}
{-M_{12}-\epsilon^*_1({\bf p})-\epsilon^*_2({\bf p})}
\nonumber\\
&+&
{\cal N}_{++}(T)\frac{
\Lambda^{(1)*}_{+}({\bf p})
\beta{\cal V}(T,\mu_g)
\phi\beta\Lambda^{(2)*}_{+}(-{\bf p})}   
{-\frac{{m^*_1}^2-{m^*_2}^2}{M_{12}}+\epsilon^*_1({\bf p})+\epsilon^*_2({\bf p})}
\nonumber\\
&+&
\left.
{\cal N}_{--}(T)\frac{
\Lambda^{(1)*}_{-}({\bf p})
\beta{\cal V}(T,\mu_g)
\phi\beta\Lambda^{(2)*}_{-}(-{\bf p})}  
{\frac{{m^*_1}^2-{m^*_2}^2}{M_{12}}+\epsilon^*_1({\bf p})+\epsilon^*_2({\bf p})}
+ \delta V^{(1)}_{\mbox{T}}\right].
\end{eqnarray}
We have used the constraint $(M^2_1-M^2_2)=({m^*_1}^2-{m^*_2}^2)$
in the cross-box diagram.
The thermal distribution functions read
\begin{eqnarray}
{\cal N}_{+-}(T)
&=&
[1-n(\epsilon^*_1({\bf p})-m^*_1)
-n(\epsilon^*_2({\bf p})-m^*_2)]
\\
{\cal N}_{-+}(T)
&=&
[1-n(\epsilon^*_1({\bf p})+m^*_1)
-n(\epsilon^*_2({\bf p})+m^*_2)]
\\
{\cal N}_{++}(T)
&=&   
[1-n(\epsilon^*_1({\bf p})-m^*_1)
-n(\epsilon^*_2({\bf p})+m^*_2)]
\\
{\cal N}_{--}(T)
&=&
[1-n(\epsilon^*_1({\bf p})+m^*_1)- 
n(\epsilon^*_2({\bf p})-m^*_2)],
\end{eqnarray}
where $n(x)=1/(\exp(x/T)+1)$ is the Fermi-Dirac distribution function.

The approximation
\begin{eqnarray}
\epsilon^*_i({\bf p})\pm M_i\approx\epsilon^*_i({\bf p})\pm m^*_I,
\label{Eq_FD}
\end{eqnarray}
is justified by the confinement condition.
In the absent of the mean fields, the confinement is satisfied by the absent 
of the time-like pole $m_1(p)+m_2(p)>M_{12}$ or with a less constraint condition $\sqrt{p^2+m^2_1(p)}+\sqrt{p^2+m^2_2(p)}>M_{12}$
where the constituent quark mass $m_i(p)$ is obtained from the gap equation.
Since, we don't solve the gap equation consistently
with the BS equation but instead we use the free quark masses, this replacement is a reasonable one to avoid the time-like pole.

The last term $\delta V^{(1)}_{\mbox{T}}$ that appears in the quasipotential equation is given by
\begin{eqnarray}
\delta V^{(1)}_{\mbox{T}}&=&
\left[
{\cal M}_{++}(T)\frac{
\Lambda^{(1)*}_{+}
({\bf p})\beta{\cal V}(T,\mu_g)\phi\beta
\Lambda^{(2)*}_{+}(-{\bf p})}
{-M_{12}+(\epsilon^*_1({\bf p})-\epsilon^*_2({\bf p}))}
+
{\cal M}_{--}(T)\frac{
\Lambda^{(1)*}_{-}
({\bf p})\beta{\cal V}(T,\mu_g)
\phi\beta\Lambda^{(2)*}_{-}(-{\bf p})}
{M_{12}+(\epsilon^*_1({\bf p})-\epsilon^*_2({\bf p}))}
\right]
\nonumber\\   
&+&
\left[
{\cal M}_{+-}(T)\frac{
\Lambda^{(1)*}_{+} 
({\bf p})\beta{\cal V}(T,\mu_g)\phi\beta
\Lambda^{(2)*}_{-}(-{\bf p})}
{+\frac{{m^*_1}^2-{m^*_2}^2}{M_{12}}-(\epsilon^*_1({\bf p})-\epsilon^*_2({\bf p}))}
+{\cal M}_{-+}(T)\frac{\Lambda^{(1)*}_{-}({\bf p})\beta{\cal V}(T,\mu_g)
\phi\beta\Lambda^{(2)*}_{+}(-{\bf p})}
{-\frac{{m^*_1}^2-{m^*_2}^2}{M_{12}}-(\epsilon^*_1({\bf p})-\epsilon^*_2({\bf p}))}
\right]
\end{eqnarray}
where
\begin{eqnarray}
{\cal M}_{++}(T)&=&
[n(\epsilon^*_1({\bf p})-m^*_1)-n(\epsilon^*_2({\bf p})+m^*_2)], \\
{\cal M}_{--}(T)&=&   
[n(\epsilon^*_1({\bf p})+m^*_1)-n(\epsilon^*_2({\bf p})-m^*_2)], \\
{\cal M}_{+-}(T)&=&
[n(\epsilon^*_1({\bf p})-m^*_1)-n(\epsilon^*_2({\bf p})-m^*_2)], \\
{\cal M}_{-+}(T)&=&
[n(\epsilon^*_1({\bf p})+m^*_1)-n(\epsilon^*_2({\bf p})+m^*_2)].
\end{eqnarray}
The contribution of $\delta V^{(1)}_{\mbox{T}}$ vanishes at zero 
temperature and we have dropped it from the numerical calculations.

The wavefunction in the plane wave components reads\cite{Zakout1}
\begin{eqnarray}
\phi^{+-}({\bf p})=\left(\begin{array}{cc}
\eta({\bf p})\frac{\sigma\cdot{\bf p}}{\epsilon^*_2({\bf p})+m^*_2}&
\eta({\bf p}) 
\\
\frac{\sigma\cdot{\bf p}}{\epsilon^*_1({\bf p})+m^*_1}
\eta({\bf p}) 
\frac{\sigma\cdot{\bf p}}{\epsilon^*_2({\bf p})+m^*_2}
& 
\frac{\sigma\cdot{\bf p}}{\epsilon^*_1({\bf p})+m^*_1}\eta({\bf p})
\end{array}\right),
\end{eqnarray}
\begin{eqnarray}
\phi^{-+}({\bf p})=\left(\begin{array}{cc}
-\frac{\sigma\cdot{\bf p}}{\epsilon^*_1({\bf p})+m^*_1}
\zeta({\bf p})& 
\frac{\sigma\cdot{\bf p}}{\epsilon^*_1({\bf p})+m^*_1}
\zeta({\bf p})
\frac{\sigma\cdot{\bf p}}{\epsilon^*_2({\bf p})+m^*_2}
\\
\zeta({\bf p})
&
-\zeta({\bf p}) 
\frac{\sigma\cdot{\bf p}}{\epsilon^*_2({\bf p})+m^*_2}
\end{array}\right),
\end{eqnarray}
\begin{eqnarray}
\phi^{++}({\bf p})=i\left(\begin{array}{cc}
\chi({\bf p})&
-\chi({\bf p})\frac{\sigma\cdot{\bf p}}{\epsilon^*_2({\bf p})+m^*_2}
\\
\frac{\sigma\cdot{\bf p}}{\epsilon^*_1({\bf p})+m^*_1}\chi({\bf p})
&
-\frac{\sigma\cdot{\bf p}}{\epsilon^*_1({\bf p})+m^*_1}
\chi({\bf p})
\frac{\sigma\cdot{\bf p}}{\epsilon^*_2({\bf p})+m^*_2}
\end{array}\right),
\end{eqnarray}
and
\begin{eqnarray}
\phi^{--}({\bf p})=i\left(\begin{array}{cc}
-\frac{\sigma\cdot{\bf p}}{\epsilon^*_1({\bf p})+m^*_1}
\psi({\bf p})
\frac{\sigma\cdot{\bf p}}{\epsilon^*_2({\bf p})+m^*_2}&
-\frac{\sigma\cdot{\bf p}}{\epsilon^*_1({\bf p})+m^*_1}
\psi({\bf p})
\\
\psi({\bf p})
\frac{\sigma\cdot{\bf p}}{\epsilon^*_2({\bf p})+m^*_2}
&
\psi({\bf p})   
\end{array}\right).
\end{eqnarray}
Hence the BS equation is decomposed into the four independent radial plane waves,
\begin{eqnarray}
\left(\begin{array}{c}
\eta({\bf p}) \\
i\chi({\bf p}) \\
i\psi({\bf p}) \\
\zeta{\bf p})
\end{array}\right)
=-{\cal G}
\left(\begin{array}{cccc}
  {\cal N}_{+-}(T){\cal V}_{+-}^{+-}
& {\cal N}_{+-}(T){\cal V}_{+-}^{++}
& {\cal N}_{+-}(T){\cal V}_{+-}^{--}
& {\cal N}_{+-}(T){\cal V}_{+-}^{-+} \\
  {\cal N}_{++}(T){\cal V}_{++}^{-+}
& {\cal N}_{++}(T){\cal V}_{++}^{++}
& {\cal N}_{++}(T){\cal V}_{++}^{--}
& {\cal N}_{++}(T){\cal V}_{++}^{-+}  \\
  {\cal N}_{--}(T){\cal V}_{--}^{+-}
& {\cal N}_{--}(T){\cal V}_{--}^{++}
& {\cal N}_{--}(T){\cal V}_{--}^{--}
& {\cal N}_{--}(T){\cal V}_{--}^{-+}   \\
  {\cal N}_{-+}(T){\cal V}_{-+}^{+-}
& {\cal N}_{-+}(T){\cal V}_{-+}^{++}
& {\cal N}_{-+}(T){\cal V}_{-+}^{--}
& {\cal N}_{-+}(T){\cal V}_{-+}^{-+}
\end{array}\right)
\left(\begin{array}{c}
\eta({\bf q}) \\
i\chi({\bf q}) \\
i\psi({\bf q}) \\
\zeta{\bf q})
\end{array}\right),
\end{eqnarray}
where the effective propagator reads
\begin{eqnarray}
{\cal G}=
\left(\begin{array}{cccc}
\frac{1}{\epsilon^*_1+\epsilon^*_2-M_{12}}& 0 & 0 & 0 \\
0& \frac{-1}{\epsilon^*_1+\epsilon^*_2-({m^*_1}^2-{m^*_2}^2)/M_{12}} & 0 & 0 \\
0& 0 & \frac{-1}{\epsilon^*_1+\epsilon^*_2+({m^*_1}^2-{m^*_2}^2)/M_{12}} & 0  \\
0& 0 & 0 & \frac{1}{\epsilon^*_1+\epsilon^*_2+M_{12}}
\end{array}\right).
\end{eqnarray}

In the present calculations, we have adopted the kernel interaction as the vector Coulomb and scalar linear confining potentials
\begin{eqnarray}
{\cal V}(T,\mu_g)=
\gamma_{\mu}\times\gamma^{\mu}V_{\mbox{V}}(T)+1\times 1 V_{\mbox{S}}(T)
\end{eqnarray}
where $V_{\mbox{V}}$ and $V_{\mbox{S}}$ represents the short and long range interaction, respectively.
The temperature dependent potential for the long-range interaction has been computed\cite{Kaczmarek1}.
It reads
\begin{eqnarray}
V_{\mbox{S}}&=&
S_0-\left[\frac{\pi}{12}-\frac{1}{6}\arctan(2rT)\right]\frac{1}{r}
\nonumber \\
&~&
+\left[a_1(T)-\frac{\pi}{3}T^2+\frac{2}{3}T^2\arctan\left(\frac{1}{2rT}\right)
\right]r+
\frac{T}{2}\ln\left[1+(2rT)^2\right].
\end{eqnarray}
The string tension
$\sigma(T)$ includes thermal corrections to the zero-temperature
string tension.
It is assumed to have the form
\begin{eqnarray}
\sigma(T)=\sigma_0\left(1-b\frac{T^2}{T^2_c}\right)^{1/2}
\end{eqnarray}
for temperatures below  $T_c$, the critical temperature of the deconfinement phase.
It is urged that the string tension does not vanish at the critical 
temperature\cite{Kaczmarek1}.
In the numerical calculations, we have taken $\sigma_0=0.22 \mbox{GeV}^2$ and $b=0.99$.
The short-range interaction is taken as
\begin{eqnarray} 
V_{\mbox{V}}(T)=-\frac{\alpha_s}{r}\exp(-\mu_{\mbox{screen}}(T) r).
\end{eqnarray}
Since the Debye screening becomes active only in the deconfined phase, 
we set $\mu_{\mbox{screen}}=0$ for $T\le T_c$.
The fitting parameters $[\alpha_s, S_0]$ are taken $[0.23,140 \mbox{MeV}]$ 
and $[0.27,-156 \mbox{MeV}]$ for charmonium and a pair of open charm mesons, respectively. 
The constituent quark masses are taken as 300 and 1300 MeV for and $m_{u,d}$ and $m_{c}$, respectively.

\section{Discussion and Conclusions}

We have derived the relativistic bound state equation for $q\overline{q}$ probed in a hot and dense matter.  
It is found that the scalar mean fields modify the constituent quark mass by 
$m_q^{\star}=m_q-\sum_S g_{qS}\sigma_S$ while the vector mean fields modify the effective bound state mass by 
$M^{\star}_{q\overline{q}}=M_{q\overline{q}}+\sum_v g_{M V}\omega_V$ 
where $M_{q\overline{q}}$ is the bound state mass without vector main fields. 
Therefore, the vector mean fields decouple from the constituent quarks in the relativistic equation while the scalar mean fields modify the relativistic bound state equation. 

The adiabatic approximation is considered in the derivation of the BS equation probed in the hot bath.
The thermal excitations of the constituent quarks are separated from the thermal excitations of the flux tube.
The flux tube thermal excitation is taken into account by adopting the central inter-quark potential derived by the lattice gauge calculations\cite{Kaczmarek1}.
However, the constituent quark thermal excitation is considered in the calculation by deriving the BS equation in the ladder and cross-box approximations using the imaginary time formalism where the summation of the Matsubara frequencies introduced in the constituent quark propagator is evaluated.
Since we have considered the free quark propagator instead of the exact gap equation in BS equation, we adopt the approximation given by Eq. \ref{Eq_FD} in the Fermi-Dirac distribution functions to justify the confinement condition. This condition ensures the confinement and avoids the thermal ionization of the bound state. We have only considered the terms of the ladder and cross-box diagrams which have zero temperature counterparts\cite{Zakout1}.
The other terms that appear in the ladder and cross-box diagrams at finite temperature 
but vanished at zero temperature are dropped from the calculations.
However, the contribution of these terms is negligible.
We are not sure if these terms are cancelled by counterparts that come from other diagrams with a thermal origin.
The BS equation with the quantum numbers $\{JLS\}$ is solved exactly. 
The full relativistic quark kinematics and their thermal excitations are considered exactly in the calculations. 
These fully relativistic kinetic effects become significant at the energies involved in 
relativistic heavy ions collisions.

We have solved exactly the BS equation where both the constituent quarks and flux tube thermal excitations are considered in the calculations.
The charmonia and charmed meson pair masses versus temperature 
are displayed in Fig. 1.
The meson masses decrease rapidly at first with temperature but when the 
temperature reaches about 50 MeV this decrease is slowed down considerably.
The mass differences between the various charmonium spectra with different quantum numbers 
decrease drastically with the temperature and become very small at high temperatures.
The hyperfine splitting for the $D$-meson decreases significantly and becomes negligible when the temperature exceeds $T\sim 100$ MeV.
Near but below the critical temperature the threshold production energy 
difference between the pair of open charm mesons and charmonium 
is roughly of the order 3-4 pions.
The mass difference between $\chi_c$ and a pair of open 
charm $D(^3S_1)$ is 
about 612  MeV while the energy difference between the $J/\psi$ (or $\eta_c$) 
and a pair of open charm $D(^3S_1)$ (or $D(^1S_0)$) is almost 
613 MeV. Just below the critical temperature, we still have
\begin{eqnarray}
M_{D\overline{D}}>> M_{\chi_c}+M_{\pi}\rightarrow M_{J/\psi}+M_{\pi}.
\end{eqnarray}
Therefore, although the $J/\psi$ can be easily excited thermally to higher 
angular momentum states, it and its excited states still can't reach 
the threshold energy production of the open charm mesons pair even 
if it absorbs a pion from the media.
In the heavy ions collisions charmonium is surrounded by the pion gas.

We have also considered the BS equation with only the flux tube thermal excitations. In this approximation the constituent quarks are assumed heavy and cannot be thermally excited. The bound state equation simply reduces to the zero temperature BS equation but with a temperature dependent potential.
In this limit, the quark and anti-quark partition functions reduce to ${\cal N}_{\pm\pm}={\cal M}_{\pm\pm}=1$.
The meson spectra masses versus temperature are displayed in Fig.2. 
The charmonia and the open charm pair masses decrease slowly with the 
temperature. However, their masses start to drop significantly just below the critical temperature.
Furthermore, the energy differences between states with different quantum numbers does not change much with the temperature. 
Therefore, the selection rules persist until almost the critical temperature. 
The mass difference between $\chi_c$ and a pair of open charm $D(^3S_1)$ is 
found to be 52, 100.1 and 192 MeV for $\chi_0, \chi_1$ and $\chi_2$, respectively.
The masses difference between $J/\psi$ and a pair of open charm $D(^3S_1)$ 
is found to be 147 MeV while it is 158 MeV for $\eta_c$ and a pair of open 
charm $D(^1S_0)$.
The minimum energy difference between the pair of open charm mesons and 
charmonium becomes of the order of a pion mass or less below the critical temperature. 
The threshold production energy is given as
\begin{eqnarray}
M_{D\overline{D}} \le
M_{\chi_c}+M_{\pi}\rightarrow M_{J/\psi}+M_{\pi}.
\end{eqnarray}
Moreover, the production energy is of the same order as the mass fitting deviation for $\chi_{c}$.
\begin{eqnarray}
M_{\chi_0}-M_{D\overline{D}}\approx 
|M^{\mbox{exp}}_{\chi_0}-M^{\mbox{fit}}_{\chi_0}|.
\end{eqnarray}
Therefore, the $J/\psi$ can be thermally excited to a higher 
orbital angular momentum state and then decays to a pair of open 
charm mesons
either spontaneously or by absorbing a pion and then decays to open charm 
mesons pair spontaneously.
This interaction raises drastically the $J/\psi$ or $\chi_c$ suppression probability.
We have compared the ground state masses for charmonia and open charm 
mesons pair  with and without constituent quark kinetic thermal 
excitations. The results are displayed in Fig. 3.
The hyperfine splitting becomes less important when the constituent quark 
thermal excitations are involved in the calculations and the threshold 
production energy to a pair of open charm mesons becomes much higher.
  
In the conclusion, our results show that there is no thermal route for the spontaneous thermal $J/\psi$ suppression in the hadronic phase when the thermal excitation of the constituent quarks is taken into account beside the temperature dependent potential derived from the lattice gauge calculations.

\acknowledgments
This work is support by
Deutsche Forschungsgemeinschaft GR 243/51-2,
Fulbright fellowship program FY2001-2002, 
and the Bergen Computational Physics Laboratory in the framework of 
the European Community's Access to Research Infrastructure
action of the Improving Human Potential Programme.

\begin{figure}
\includegraphics{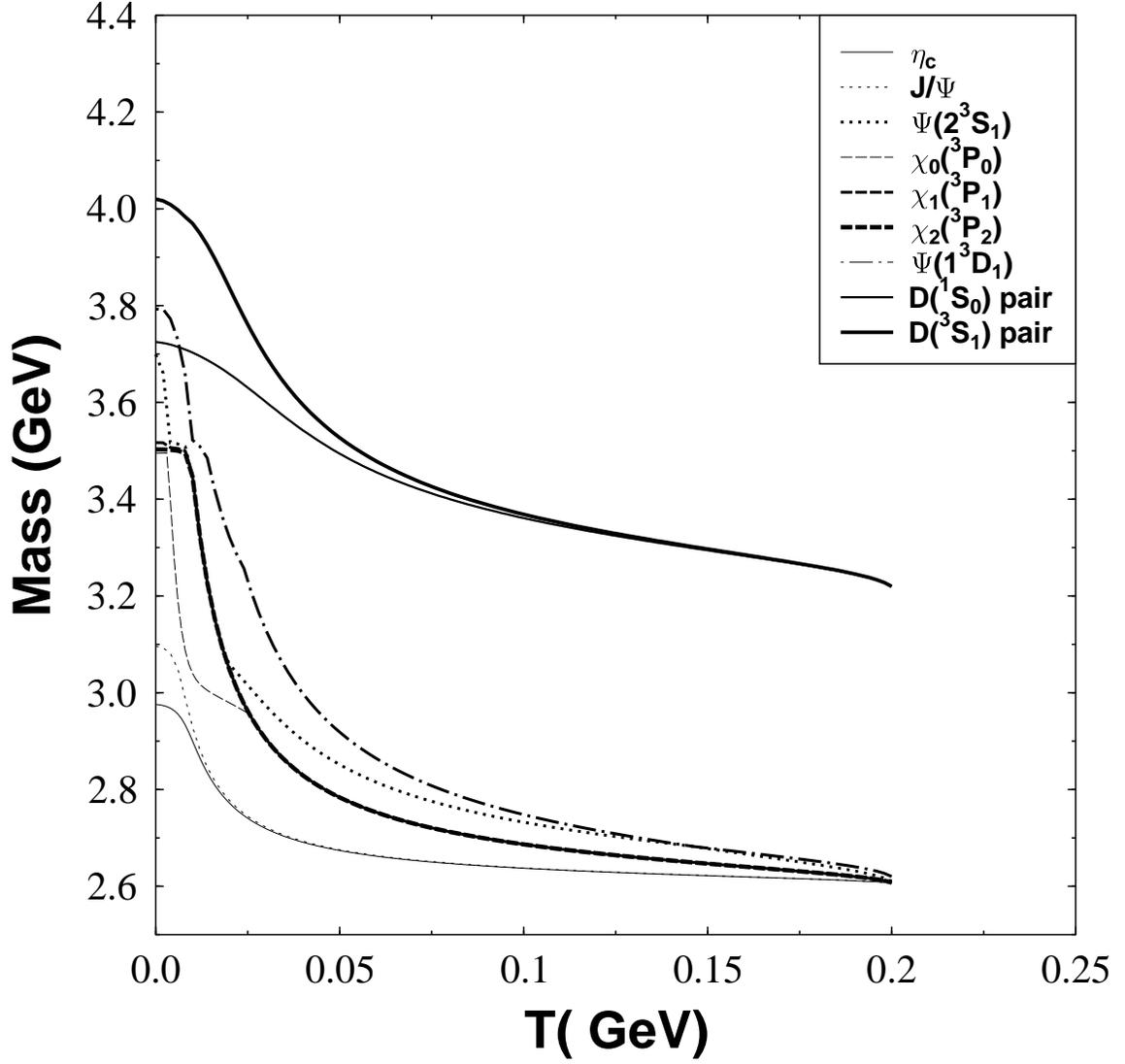}%
\caption{\label{fig1:psart}
Charmonium and a pair of open charm mesons $D\overline{D}$ spectra versus 
temperature with thermal flux tube and quark excitations.}
\end{figure}
\begin{figure}
\includegraphics{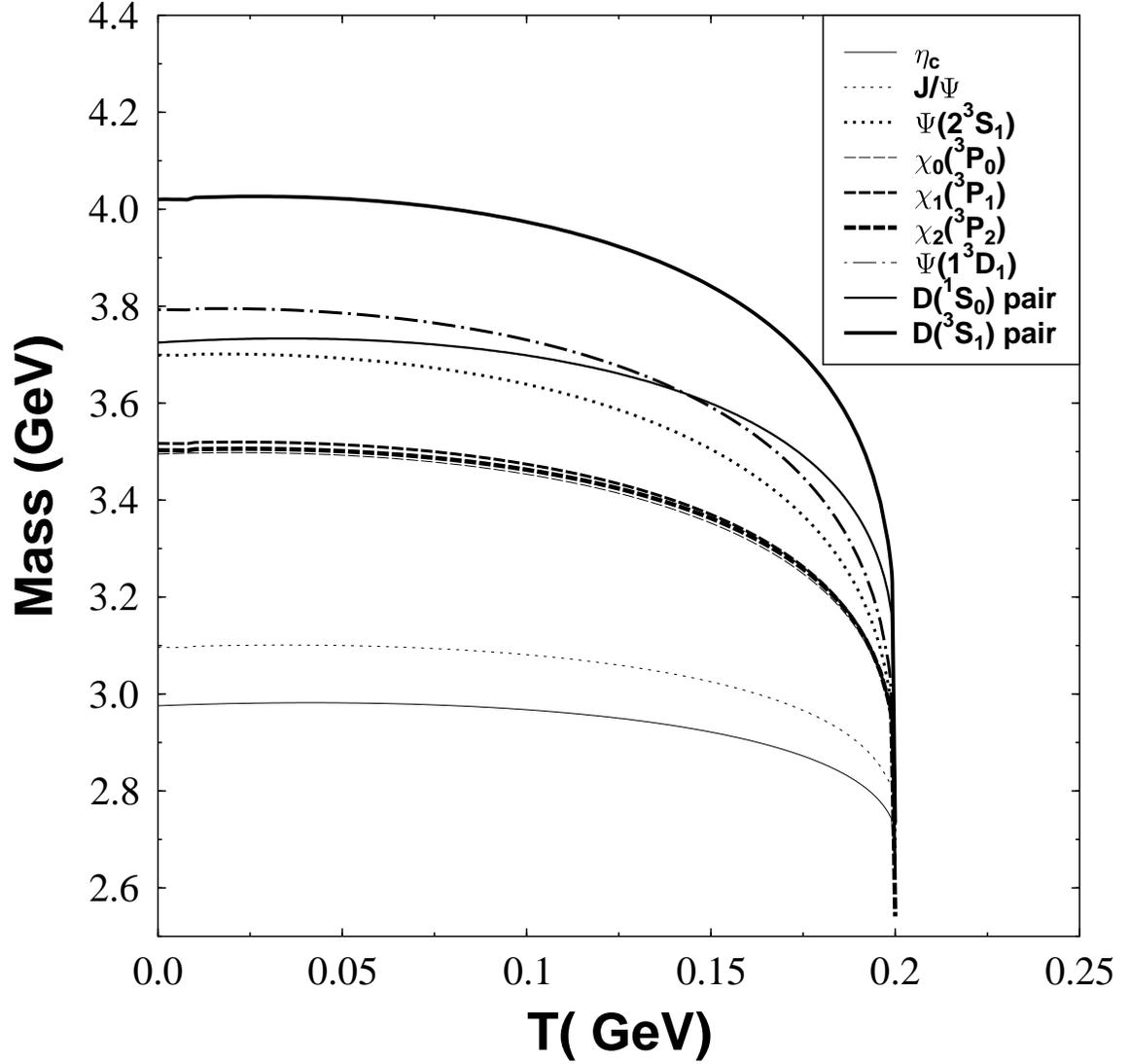}%
\caption{\label{fig2:psart}
Charmonium and a pair of open charm mesons $D\overline{D}$ spectra versus 
temperature
with only thermal flux tube excitation.}
\end{figure}
\begin{figure} 
\includegraphics{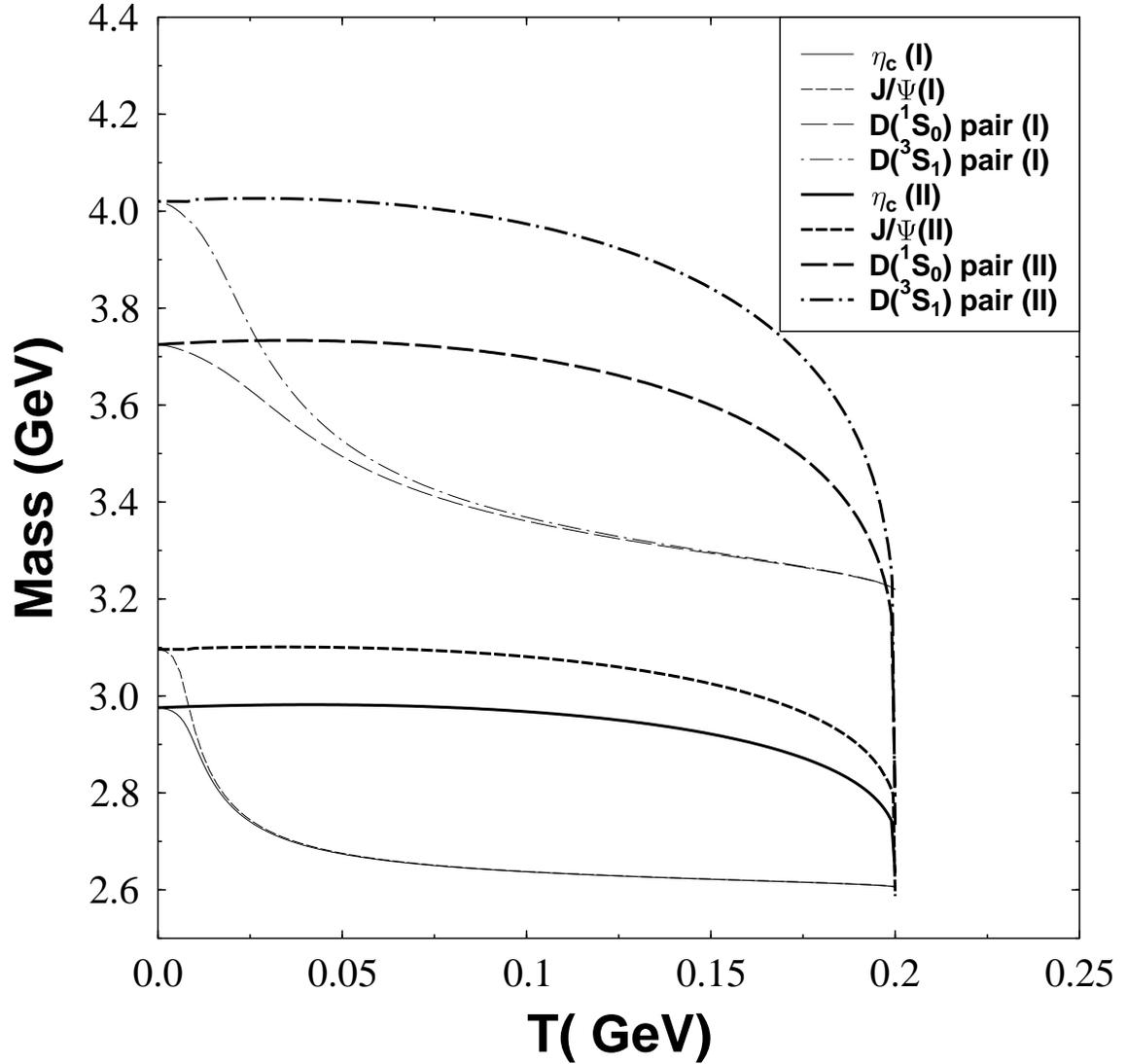}%
\caption{\label{fig3:epsart}
Comparison between the spectra of charmonium and a pair of open charm 
mesons production with the thermal flux tube excitation and the thermal 
quark and flux tube excitations.}
\end{figure}     
\end{document}